%% file: main.tex
\documentclass[]{spie}  %

\usepackage{aas_macros}
\usepackage{amsmath,amsfonts,amssymb}
\usepackage{color}
\usepackage{caption}
\usepackage{subcaption}
\usepackage{graphicx}
\usepackage[colorlinks=true, allcolors=blue]{hyperref}
\usepackage{xspace}

\title{The Japanese Vision for the Black Hole Explorer Mission}

\input{authorlist}

\authorinfo{Further author information: (Send correspondence to K.A.)\\K.A.: E-mail: kakiyama@mit.edu, Telephone: 1 617 715 5579}

\def\M87{M87$^*$\xspace}
\def\m87{M87$^*$\xspace}
\def\sgra{Sgr\,A$^*$\xspace}
 
\begin{document} 
\maketitle

\begin{abstract}
The Black Hole Explorer (BHEX) is a next-generation space very long baseline interferometry (VLBI) mission concept that will extend the ground-based millimeter/submillimeter arrays into space. 
The mission, closely aligned with the science priorities of the Japanese VLBI community, involves an active engagement of this community in the development of the mission, resulting in the formation of the Black Hole Explorer Japan Consortium.
Here we present the current Japanese vision for the mission, ranging from scientific objectives to instrumentation.
The Consortium anticipates a wide range of scientific investigations, from diverse black hole physics and astrophysics studied through the primary VLBI mode, to the molecular universe explored via a potential single-dish observation mode in the previously unexplored 50-70\,GHz band that would make BHEX the highest-sensitivity explorer ever of molecular oxygen. 
A potential major contribution for the onboard instrument involves supplying essential elements for its high-sensitivity dual-band receiving system, which includes a broadband 300\,GHz SIS mixer and a space-certified multi-stage 4.5K cryocooler akin to those used in the Hitomi and XRISM satellites by the Japan Aerospace Exploration Agency. 
Additionally, the Consortium explores enhancing and supporting BHEX operations through the use of millimeter/submillimeter facilities developed by the National Astronomical Observatory of Japan, coupled with a network of laser communication stations operated by the National Institute of Information and Communication Technology.
\end{abstract}


\section{INTRODUCTION}
\label{sec:intro}  %
Black holes are among the most profound and fundamental prediction of Einstein's theory of general relativity (GR)\cite{Einstein_1915, Schwarzschild_1916, Kerr_1963}. 
The production of black holes is believed to be generic in the universe\cite{Oppenheimer_1939, Penrose_1965}, making them unique probes to explore gravitational physics in strong-field regime \cite{EHTC2017SgrAPaper6,Ayzenberg_2023}. 
Beyond providing an important probe for fundamental physics, black holes are now widely known to play a central role in a wide variety of astrophysical phenomena with their unique capabilities of efficiently extracting and releasing energies from the accreting material \cite{Lynden-Bell_1969, Penrose_1971, Yuan_2014}.
On the heaviest side of the known populations, nearly every galaxy hosts a supermassive black hole (SMBH) in its center \cite{Richstone_1998}, which may play a role in its evolution\cite{Magorrian_1998,Fabian_2012,Kormendy_2013}.
In particular, a large amount of mass accretion to the central SMBH powers a luminous active galactic nucleus (AGN)\cite{Lynden-Bell_1969, Urry_Padovani_1995}, where rapidly spinning supermassive black holes are thought to launch and power relativistic jets\cite{Blandford_Znajek_1977} that often extend beyond their host galaxies, ultimately shaping evolution on cosmological scales\cite{Blandford_2019}.

Very long baseline interferometry (VLBI) of radio astronomy has been a central tool providing the closest look at the vicinity of black holes, known as the most compact objects in the universe for their given masses, leveraging its uniquely high spatial resolutions. 
With the advent of ground-based global millimeter/submillimeter VLBI arrays, the innermost regions near the vicinity of the black hole's event horizon are now accessible for studies with direct imaging.
The Event Horizon Telescope (EHT)\cite{EHTC2017M87Paper2}, 
operating at 0.87 and 1.3\,mm wavelengths (345 and 230\,GHz, respectively, in frequencies), has enabled the transformative science of studying black holes on event horizon scales. 
EHT has provided the first-ever spatially resolved images of black hole shadows captured for two supermassive black holes:
Messier 87$^*$ (\m87) at the heart of the nearby radio galaxy M87\cite{EHTC2017M87Paper1,EHTC2017M87Paper4,EHTC2017M87Paper7,EHTC2017M87Paper9,EHTC2018M87Paper1},
and Sagittarius A$^{*}$ (\sgra) at the center of our Milky Way galaxy\cite{EHTC2017SgrAPaper1,EHTC2017SgrAPaper3,EHTC2017SgrAPaper7}.
The Global Millimeter VLBI Array (GMVA), 
operating at 3\,mm wavelength (90\,GHz), further provided the first direct image of a black hole accretion flow around \m87 connecting the base of the powerful jet up to EHT images of the horizon-scale emission\cite{Lu_2023}. 
Both EHT and GMVA have provided the innermost structures of many AGN jets\cite{Boccardi_2016, Janssen_2021, Okino_2022, Zhao_2022, Issaoun_2022, Jorstad_2023, Paraschos_2024}, resolving the key region where its magnetized plasma flow is being shaped into a powerful collimated flow, accelerated to a relativistic speed, and potentially further heating particles that will produce high-energy cosmic rays and electromagnetic waves \cite{Boccardi_2017, Blandford_2019}.

However, those scientific breakthroughs simultaneously revealed the severe limitations of their angular resolutions, restricted by the diameter of the Earth and severe atmospheric effects that prevent from forming an array at a higher frequency. 
At the EHT resolution, while the structures controlling inflow and outflow around black hole start to be resolved \cite{EHTC2017M87Paper5, EHTC2017M87Paper8, EHTC2017SgrAPaper5, EHTC2017SgrAPaper8}, the ability to constrain the black hole spacetime or its role in powering black hole jets is severely limited by the lack of angular resolutions \cite{EHTC2017M87Paper6, EHTC2017SgrAPaper6, Tiede_2022, Ayzenberg_2023}. 
The angular resolution limitations of the Earth-bound VLBI prevent it from producing event-horizon-scale images of black holes and their shadows, except for the two currently observable sources, \m87 and \sgra\cite{Johannsen_2012, Pesce_2022, Ramakrishnan_2023}.
Those science questions, critical to comprehending the black holes in cosmic and fundamental physical contexts, require an extension of global millimeter/submillimeter VLBI arrays into space. 

The Black Hole Explorer (BHEX)\cite{BHEX_Johnson_2024}\footnote{\url{https://www.blackholeexplorer.org/}, accessed on May 29, 2024} is a next-generation space VLBI mission concept, being developed for the Astrophysics Small Explorer (SMEX) program of the National Astronautics and Space Administration (NASA). 
BHEX aims primarily at the first direct detection of a black hole’s ''photon ring'' comprising rays of light that have orbited the black hole before escaping. The photon ring traces a narrow region of space just outside the black hole’s event horizon, which is a unique probe for its spacetime and enables direct measurements of a black hole's spin as well as a precise test of general relativity\cite{Johnson_2020, Gralla_2020, Palumbo_2023, Lupsasca_2024, BHEX_Lupsasca_2024, BHEX_Galison_2024, BHEX_Kawashima_2024}. 
BHEX further enables the demographic studies of the event-horizon-scale properties for dozens of additional SMBHs, providing crucial insights into the processes that drive their creation and growth. 
Additionally, BHEX will link these SMBHs with their relativistic jets, shedding light on the energy mechanisms behind the universe's most luminous and effective engines.

To address those scientific questions, BHEX explores the universe at the highest angular resolution ever in the history of astronomy, surpassing what is possible with the terrestial telescopes alone, by extending the existing ground-based millimeter/submillimeter VLBI networks such as EHT and GMVA to space.
BHEX is currently designed to be a dual-band instrument, capable of simultaneously capturing signals in two frequency bands\cite{BHEX_Johnson_2024, BHEX_Marrone_2024} to utilize the frequency phase transfer technique \cite{Asaki_1996, Dodson_2009, Rioja_2011}. 
The operating frequency ranges for the receiver are set at 100 GHz (80-106\,GHz) and 300 GHz (240-320\,GHz)\cite{BHEX_Marrone_2024, BHEX_Tong_2024}, overlapping the frequency bands used by GMVA and EHT, respectively.
BHEX is designed to be a sensitive instrument that achieves the baseline fringe sensitivity of the $\sim 1$\,mJy, $\sim$ 100\,times better than that of current GMVA that has already detected more than 160 compact sources in the sky\cite{Nair_2019}. 
With its sensitivity, BHEX provides an unprecedented capability of observing hundreds of extragalactic sources at a spatial resolution of $\sim 5$\,${\rm \mu}$as.

The Japanese community has been an important contributor to shaping the BHEX mission concept, from science goals to instrument design. 
BHEX is a natural extension of ongoing VLBI research in Japan, closely aligning with the community's strategic priorities identified for the 2030s. 
The mission offers a distinct opportunity to meet many key scientific objectives in the research of AGN and SMBH recognized by the community. 
The collaboration between BHEX and the Japanese community resulted in the establishment of the Black Hole Explorer (BHEX) Japan Consortium, which seeks to advance BHEX as a collaborative space program between the United States and Japan. 
Here, we present the current Japanese vision for BHEX. 
We first describe the BHEX Japan Consortium along with its community context in \autoref{sec:bhex_japan}. For the community vision, we outline the science drivers for BHEX envisioned by the Japanese community in \autoref{sec:sci_cases}, and potential technical contributions as a key cost-share partner of the mission in \autoref{sec:tech_cases}.
We will summarize our vision in \autoref{sec:summary}.
This paper is part of a series of articles that describe the current concept of the BHEX mission \cite{BHEX_Johnson_2024, BHEX_Marrone_2024, BHEX_Peretz_2024, BHEX_Lupsasca_2024, BHEX_Galison_2024, BHEX_Issaoun_2024, BHEX_Kawashima_2024, BHEX_Tomio_2024, BHEX_Wang_2024, BHEX_Sridharan_2024, BHEX_Rana_2024, BHEX_Tong_2024, BHEX_Srinivasan_2024}. 

\section{The Black Hole Explorer Japan Consortium}
\label{sec:bhex_japan}

\subsection{BHEX in the Japanese VLBI Context}
\label{sec:bhex_japan_context}
Japan has been a leading country for high-resolution studies of SMBHs, AGNs, and their jets with VLBI. 
The community is well known for the VLBI Space Observatory Programme (VSOP) of the Japan Aerospace Exploration Agency (JAXA) \cite{Hirabayashi_1998}, the first successful astronomy space VLBI mission that further motivated its planned (but halted) successor VSOP-2 program\cite{Hirabayashi_2004} and the successful Russian RadioAstron space VLBI program\cite{Kardashev_2013}. 
The community has conducted a wide variety of VLBI studies with its own ground-based VLBI instruments, including the VLBI Exploration for Radio Astrometry (VERA)\cite{Honma_2000} and the greater Japanese VLBI Network (JVN)\cite{Doi_2006}, which are now extended to the East Asian VLBI Network (EAVN)\cite{Akiyama_2022} and larger global VLBI arrays, including the East Asia to Italy Nearly Global (EATING) VLBI program\cite{Giovannini_2023} and the Global VLBI Alliance (GVA) Program\cite{Colomer_2023}. 

The Japanese community has also been known as a major science and technical contributor to EHT since its early era\cite{Akiyama_2015, Asada_2017} and the ALMA Phasing Project\cite{Matthews_2018} providing the key VLBI capability to the Atacama Large Millimeter/Submillimeter Array (ALMA) for EHT and GMVA. 
Dozens of scientists in the EHT Japan Group have been playing key roles in its groundbreaking science results and the overall operations of EHT through the international EHT collaboration. 

Such a diverse nature of VLBI programs in Japan, not limited to astronomy, has been supported by a broad interdisciplinary community represented by the Japan VLBI Consortium\footnote{\url{https://www2.nict.go.jp/sts/stmg/vcon/index-e.html}, accessed on May 29, 2024} established in 1990, involving 135 scientists as of December 2023. 
In 2020, the Japan VLBI Consortium initiated a large community-wide survey for the community vision toward the 2030s. 
As an outcome of the survey, many BHEX science cases were identified as key science goals in the research area of SMBHs, AGNs, and their jets in their community roadmap toward the 2030s, published in 2021 by the Consortium's Future Planning Working Group\cite{Akahori_2021}.

The BHEX mission concept and its vision are aligned with the community's priorities for the 2030s, fully leveraging the heritage from past space and ground VLBI programs in Japan.
BHEX provides a unique opportunity to accomplish many of the community's key science goals in the area of AGN and SMBH studies.
Given strong synergies between BHEX and the community vision, the BHEX Japan Consortium was formed to join the development of the international BHEX mission concept and to explore the possibility of making BHEX a US-Japan international space program. 

\subsection{The Black Hole Explorer Japan Consortium}
\label{sec:bhex_japan_consortium}

The BHEX Japan Consortium was formed in September 2023 to investigate the scientific aspects of the mission, as well as potential contributions to the BHEX instrument and its operations. 
By May 2024, the Consortium had grown to include over 60 scientists from more than 25 institutions. 
The BHEX concept studies in Japan leverage the wide range of expertise of the participants, encompassing observational radio astronomy using both single dishes and radio interferometry not limited to VLBI, as well as astronomy instrumentation and technology development, theoretical physics/astrophysics, and optical laser communication. 
The Consortium and its subgroups consistently convene to advance the concept of Japanese contributions, maintaining communication with the broader US-led BHEX collaboration.

The Consortium has three science working groups, each tasked with developing specific science cases and mission requirements: the General Relativity, Accretion, and Jet Launching Working Group focusing on event-horizon-scale physics and astrophysics, the AGN Working Group addressing wider studies of SMBHs, and the Single Dish Working Group investigating scientific possibilities in a potential single-dish mode.
The ongoing scientific vision formulated by these working groups will be detailed in \autoref{sec:sci_cases}.
For the potential technology and instrument contributions, the Consortium assembles experts from past Japanese space VLBI missions, ground-based radio asronomy instruments not limited to VLBI, and laser communication, which is a new key technology of the BHEX mission.
Four key techonology areas have been identified for the potential Japanse contributions to the BHEX mission, as described in \autoref{sec:tech_cases}.

\section{Scientific Opportunities with BHEX}
\label{sec:sci_cases}
The mission-defining science goal of BHEX is the discovery of a black hole photon ring, produced by light that has orbited the black hole before escapting\cite{BHEX_Johnson_2024}. The photon ring is a {\it universal} signature expected for black holes whenever the emission is optically thin. The surrounding light-emitting material has negligible effects on its appearance shaped by the trajectory of light along with the geodestics of the black hole spacetime. The resolved measurement of a photon ring allows a unique exploration of the black hole spacetime at an unprecedent precision not possible from the ground.
The mission aims to discover and precisely measure the black hole photon rings created by \m87 and \sgra, which has been identified as a key science objective of the Japanese VLBI Consortium's roadmap for the 2030s\cite{Akahori_2021}.

Based on the requirements of photon ring science, the current operational concept \cite{BHEX_Issaoun_2024} has two years of mission lifetime from launch to decommissioning, of which each year will host a pair of three-month observing windows for the photon ring science opportunity, each for \m87 and \sgra, from January to March and from June to August, respectively. 
In those observing windows, the mission aims to make repeated observations on those primary targets at a cadence of every $\sim$3 days. 
In aggregate, about a year of the observing time would be available for the secondary science cases, supposed to be spent for other targets. 
Besides, even during the duration of the photon ring science operations, there may be the possibility of filler-type observations to maximize the duty cycles and science output of the mission operations. 

Besides the primary photon ring science case that has drawn a decade of community attention in Japan, the BHEX Japan Consortium has explored various science opportunities uniquely enabled by BHEX both for the two primary targets \m87 and \sgra and for various secondary target sources within the three science working groups described in \autoref{sec:bhex_japan}. 
Here, we describe the current outlook and snapshot of the community vision for BHEX science. 
We first describe the potential science opportunities for the event-horizon-scale physics and astrophysics near the black holes in \autoref{sec:sci-horizon} and for a wide variety of AGNs in \autoref{sec:sci-agn}, enabled by BHEX VLBI operations. 
We further describe the scientific opportunities that can be carried out {\it with the BHEX sattelite alone} using a potential single-dish observing mode in \autoref{sec:sci-sigle-dish}, given the nature of the mission that requires millimeter/submillimeter VLBI stations of which availability may be significantly constrained due to the weather conditions and operational plans of partner observatories.

\subsection{Science Opportunities for Event Horizon Scale Physics and Astrophysics}
\label{sec:sci-horizon}
A central focus of the BHEX science lies in its capability of resolving the emission near the black hole event horizon.
For the discovery of the photon ring, BHEX will achieve angular resolution on the order of $\sim 0.1$ Schwartzchild radii.
BHEX is also capable of resolving event-horizon-scale emission for dozens of nearby SMBHs at a resolution of a few Schwartzchild radii, comparable to the ground EHT observations of \m87 and \sgra.
Significant expansion of horizon-scale targets, currently limited to only two sources, will thereby facilitate demographic studies of these environments and their influence on the generation of relativistic jets.
With those extraordinary capabilities, BHEX will probe the black hole spacetime as well as the extremely energetic plasma properties including the magnetic fields of the accretion flows and relativistic jets. 
These science goals are identified as key science goals in the Japanse VLBI community for the next decade\cite{Akahori_2021}.

We here summarize the science cases to be addressed by BHEX, some described in the last community survey\cite{Akahori_2021}, which will be complementary to the above science opportunities with BHEX (see the companion proceeding\cite{BHEX_Kawashima_2024} in detail).
\begin{itemize}
    \item {\bf Spin Measurement with Crescent-like Shadow}\\
    The “crescent-like shadow”, as we refer to the dark region formed between the photon ring and the direct image of the accretion flow\cite{kawashima_2019} in slightly brighter states. As the magnitude of the black hole spin increases, the width of the crescent-like shadow becomes thicker, since the center of the photon ring shifts more prominently attributed to the frame-dragging effect (i.e., the effect of the black hole). The width of the crescent-like shadow can constrain the black hole spin. This feature, clearly resolved only from space for \m87 and \sgra, will be complementary to the direct and precise measurement of the photon ring.
    \item {\bf Capturing the Signatures of Mass Loading in the Jet Launching Region}\\
    One of the main questions in jet launching is how and where the material is supplied in the jet, so called mass loading. In particular, plasma should be injected into the highly magnetized region if the jet is powered by the extraction of energy with black hole spin through the Blandford-Znajek process\cite{Blandford_Znajek_1977}. It is proposed that the multiple ring-like feature will appear with diameters of $\sim 60\,\mu {\rm as}$ around \m87\cite{kawashima_2021,ogihara_2024} if its jet is powered by the rapidly spinning black hole and the plasma is injected on and near the separation surfaces connecting inflows and outflows inside the highly magnetized jets. Those signatures, a unique probe for the mass loading mechanism for the jet, contaminated with other emission features at the ground EHT resolution and only accessible at the BHEX resolution.
    \item {\bf Exploring the black hole magnetosphere through polarimetry} \\
    Synchrotron polarization is affected by the Faraday effect near a black hole; linear polarization (LP) vectors are scrambled by Faraday rotation, and circular polarization (CP)  components are increased by Faraday conversion. As a result, they show different distributions on the ring (LP-CP separation)\cite{tsunetoe_2022,Tsunetoe_2022b}, which enable us to survey the magnetic fields, electron temperature near black hole, and possibly jet launching mechanism. 
    With respect to the particle acceleration in jets near a black hole, it is known that opaque nonthermal electrons produce 90°-flipped LP vectors and oppositely handed CPs, compared to optically thin plasmas. In nonthermal jet images, this polarization flipping appears on the photon ring\cite{tsunetoe_2024}. We can expect to investigate nonthermal particles through BHEX observations of \m87, \sgra and newly resolved SMBHs.
    \item {\bf Test of Gravitational Physics Theories} \\
    The photon ring and overall BHEX images of \m87 and \sgra will significantly improve the capability of exploring the deviation from the Einstein’s theory. For example, Refs. \citenum{mizuno_2018} and \citenum{fromm_2021} has been studied the comparison of the shadow images of Kerr black hole and the dilaton black hole. It was found that these can be distinguished by using the space-VLBI. Recently, the possible Rezzolla-Zhidenko metric has also been explored. These are being carried out by international collaborations.
\end{itemize}
With respect to the relation to \autoref{sec:sci-agn}, these studies complement each other: horizon-scale studies will provide an inner boundary condition of the AGN jets and vice versa.

\subsection{Scientific Opportunities for Active Galactic Nuclei}
\label{sec:sci-agn}
Given the highest angular resolution enabled by BHEX, we have identified three scientific opportunities that can be carried out with the BHEX mission, along with the community's priorities\cite{Akahori_2021}: 1) understanding how relativistic jets are produced in the vicinity of SHBHs, 2) understanding how relativistic jets accelerate particles, and 3) constraining the most accurate masses of SMBHs and the Hubble constant.
In the following, we briefly describe our current vision for each of the three science goals. 

\subsubsection{Addressing the Universarily of the Jet Launching Mechanisms}
Understanding the launching jets from SMBHs in AGNs remains a challenge in astrophysics, which is a key science goal of the BHEX mission\cite{BHEX_Johnson_2024}.
EHT has enabled studies from the black hole event horizon to the terminal end of the relativistic jet in M87\cite{EHTMWL_2021, EHTMWL_2024}, a representative of low power jets seen in Fanaroff-Riley I (FR I) radio galaxies.
BHEX will provide the first-ever direct measurement of a black hole spin for \m87 from the shape of its photon ring\cite{Johnson_2020,Gralla_2020}, which will address the role of the energy extration by the black hole spin in the production of a powerful relativistic jet\cite{Blandford_Znajek_1977}.
The newly resolved event-horizon-scale emissions from dozens of nearby SMBHs (\autoref{sec:sci-horizon}) will further allow demographic studies connecting the black hole event horizon with the jet as being explored for M87 with the current EHT.

The high angular resolution and sensitivity achieved with BHEX further enable to address the processes of the jet launching and evolution in a wide variety of AGNs. 
Given the fact that M87 and those new horizon-scale targets have low-power jets from low luminosity AGNs and less-powerful FR I galaxies thought to be powered by the same class of hot accretion flows\cite{Yuan_2014}, expanding classes of targets are imperative to understand the universality of those processes and ultimately the cosmic roles of the black holes and jets.
In this context, the BHEX Japan Cornsortium currently identify the following target sources that have substantially different accretion rates, activities, and nucleus enviroments, as their high-priority targets. 
These targets, extensively researched by the Japanese community, encompass a high-power jet from an FR II radio galaxy at the completely opposite end of M87, a nascent jet source developing through interactions with the interstellar medium, relativistic jets from rapidly evolving supermassive black holes, and relativistic jets from a mysterious category of AGNs whose positions are still debated in the existing AGN unification framework. 

\paragraph{Cygnus A}
Cygnus A, an archetypal high-power FR II radio galaxy, has been identified as the primary target to address the jet launching mechanism for the FR-II type jet, providing a compelling contrast to M87. 
Previous observations \cite{Boccardi_2016} revealed intriguing features at Cygnus A's jet base, including an unusually wide jet base width, suggesting the presence of additional components like an accretion disk or wind. 
To investigate the origin of the excess emission observed at Cygnus A's jet base, we propose BHEX imaging observations at a scale of approximately 100 Schwarzschild radii ($R_{s}$). 
These high-resolution images will allow us to discern the spatial distribution of emission and probe the presence of associated structures, such as an accretion disk or wind component. 
By obtaining spectral index information of Cygnus A's jet base through BHEX imaging, we anticipate gaining insights into the physical processes responsible for the observed excess emission, specifically whether it originates from the surrounding wind or the accretion disk, thus shedding light on the jet launching mechanism in high-power AGN jets.

\paragraph{3C 84}
In a scenario where the structures of relativistic jets grow in a self-similar way, the evolution of radio-loud AGN depends on its linear size. 
At the smallest end, the sources smaller than a few tens of parsecs are known as the youngest populations of AGN jets, which are either newborn or produced by episodic short-lived AGN activities. 
Those sources offer unique opportunities to study the evolutional process of the relativistic jets and their affects on interstellar environments in their host galaxies. 
3C 84 (NGC 1275), one of the brightest nearby AGNs at radio wavelengths, is an exceptional source among known young radio sources. 
A new episode of flaring jet activity, which started around 2005, has been forming a compact jet + lobe structure that propagates through the circumnuclear environment at $\sim$1\,pc. In the past decade, VLBI observations revealed that this “restarted'' jet has been showing dramatic changes in its motion, as well as the entire morphology \cite{Nagai_2014, Nagai_2017, Kino_2017, Kino_2021}. 
Such events can be interpreted as the interaction with the ambient gas in the circumnuclear region. 
Polarized emission from the jet also provides a clue to measure the electron density and magnetic field of the ambient gas through the measurement of Faraday rotation\cite{Nagai_2017}. 
Thanks to its brightness, 3C 84 can be easily detected even with space-ground baselines, and therefore this source can be an excellent target for the BHEX mission. 
The ultra high angular resolution of the BHEX would be promising to reveal detailed time variation of the jet structure even at the inner region. 
BHEX might also be useful to pinpoint the location of clouds that are responsible for the absorption in the HCO+ and HCN spectra found by ALMA observations \cite{Nagai_2019}. This will give a hint for the role of cold accretion in terms of AGN fueling.

\paragraph{Narrow Line Seyfert I Galaxies}
Narrow Line Seyfert I (NLSy1) galaxies are another class of AGNs, believed to be powered by relatively small supermassive black holes ($10^6-10^8$\,M$_\odot$) at the high accretion rates that almost reach the Eddington accretion rate \cite{Boller_1996}.
Nearly $\sim 7$\% of NLSy1s are radio-loud\cite{Zhou_2006}, and therefore have relativistic jets, which offer a unique laboratory for studying the physics and roles of jets ejected from rapidly evolving supermassive black holes.
Given the nature of small black hole masses, the current ground-based VLBI arrays can only access the outer areas compared to those observed in other nearby AGNs. 
The extreme angular resolution offered by BHEX observations would provide the unique opportunity to study jet-launching mechanisms and circumjet nuclear environments.

\paragraph{Broad Absorption Line Quasars}
A significant proportion of quasars exhibit blue-shifted absorption troughs in their rest-frame ultraviolet spectra, known as broad absorption lines (BALs), which result from broad resonance lines\cite{Weymann_1991}. The origin of BALs is thought to be related to the accretion disk wind, but there remains uncertainty as to whether the observed ratio of BAL to non-BAL quasars is due to orientation. Although the radio properties of BAL quasars are statistically consistent with the picture that they are primarily observed at a large viewing angle relative to the jet axis\cite{DiPompeo_2011,Hayashi_2024}, a subset of objects is considered to be observed from a pole-on view, named polar BAL quasars\cite{Ghosh_2007}. Thus far, some of these have been observed using ground-based facilities at centimeter wavelengths, revealing a one-sided jet or significant variability \cite{Bruni_2013,Hayashi_2013,Kunert-Bajraszewska_2015,Ceglowski_2015,Ceglowski_2017}. Despite these efforts, the sub-pc-scale properties of these objects remain to be revealed because most BAL quasars have been found at high redshifts where ultraviolet absorption lines fall within the optical band. Utilizing extremely long baselines with BHEX will resolve the spatial structure of polar BAL quasars at high redshift. The origin of polar BAL quasars and their jet activity can be elucidated by comparing their radio characteristics with those of typical flat-spectrum radio quasars and blazars.

\subsubsection{Understanding the Roles of AGN Jets as a Cosmic Particle Accerelator}
With the advent of sensitive $\gamma$-ray telescopes such as the Fermi Gamma-ray Space Telescope and ground-based Cherenkov Telescopes such as VERITUS, MAGIC and HESS, relativistic jets in AGNs were identified as the most numerous populations of $\gamma$-ray sources on the extragalactic sky\cite{Madejski_2016}. 
A major emission mechanism of $\gamma$-ray emission from AGN jets, some reaching very high-energy regimes beyond TeV, is considered to be the inverse-Compton scattering process involving relativistic electrons particle-accerelated within AGN jets.
Now, AGN jets are widely believed to be the most promising class of cosmic-ray accerelators, potentially up to the ultra-high-energy regime beyond $10^{18}$\,eV\cite{Rieger_2022}.
The significant roles of AGN jets as a cosmic particle accelerator have been strongly supported by the recent detections of high-enery neutrino reaching $100$\,TeV by the IceCube Neutrino Observatory on the two AGN sources: the blazar TXS 0506+056\cite{IceCube_2018} and the Seyfert galaxy NGC 1068\cite{IceCube_2022}.

Blazars are a class of AGNs that dominate the extragalactic $\gamma$-ray sky\cite{Madejski_2016}. 
The electromagnetic emission from blazars, the relativistic jets of which point to our direction \cite{Urry_Padovani_1995}, is significantly magnified by the relativistic beaming effects, offering unique opportunities to study the emission and particle accerelation processes through observations across the entire electromagnetic spectrum.
Blazars, because of this nature, will be a major class of target sources observable with BHEX. 
With extensive multi-wavelength studies of blazars, broadband blazar emissions are believed to be produced in compact emission zones often called ``blazar zones''.
One of the main focus on blazar studies has been elucidating the characteristics of the blazar zones across the wide populations of $\gamma$-ray blazars detected with Fermi.

A fundamental question for blazar zones is how and where those zones are produced within AGN jets, which directly addresses the sites of particle acceleration and high-energy emissions.
In the case of $\gamma$-ray blazars, which typically exhibit variability on the scale of days, the blazar zones are anticipated to extend to approximately 0.01\,pc\cite{Hirashita2016}, which need extremely high angular resolutions. 
At the current EHT resolutions, such compact scales are only accesible to a limited number of nearby representative TeV blazars (i.e. a single class of them), such as Mrk 501 and Mrk 421.
Futhermore, many blazars display $\gamma$-ray flares lasting from hours down to minutes\cite{Aharonian+07a}, suggesting the existence of smaller emission zones. 

BHEX provides an unprecedented opportunity to image and potentially resolve the blazar zones with its extremely high angular resolution, not accessible from the current terrestrial VLBI.
Full polarization images of the blazar zones and their surrounding area will address the roles of the magnetic fields in the particle accerlations and further help to understand the origin of complex polarization behaviors revealed by X-ray polarimetry with the Imaging X-ray Polarimetry Explorer (IXPE) sattelite\cite{Liodakis2022Natur,DiGesu2023NatAs}.
The extensive baselines of BHEX without imaging will offer the potential to investigate rapid time variability akin to $\gamma$-ray flares in nearby TeV blazars, owing to their highly core-dominated nature. 
Coordinated, simultaneous multiwavelength observations with high temporal resolution will be imperative for pinpointing the blazar zone within the jet, and potentially identifying sites of neutrino production\cite{Murase+18} in the era of multimessenger astronomy.

\subsubsection{Constraining Masses of SMBHs, Hubble Constant and Angular Momentum Transfer}
The 22 GHz water megamasers have been found within 0.1-1\,pc from the central engines of AGNs, on the order of 10$^4$--10$^5$ $R_{s}$,  by ground VLBI observations, which demonstrated that some megamasers trace thin disks that could be extensions of warmer inner accretion disks surrounding SMBHs\cite{Miyoshi_1995, Herrnstein_1999, Greenhill_2003}. 
Recent high-resolution images of the water megamaser at sub-millimeter band show that the 321 GHz water maser seems to probe the innermost part ($<$ 0.1 pc) of the disk as such high-frequency masers occur at regions with higher temperatures and densities\cite{Kameno_2024}. A few fundamental scientific questions that can be addressed by observations of the 321 GHz water megamasers using BHEX are broken down below.
\begin{itemize}
\item {Masses of SMBHs: The masses of SMBHs can be measured with less uncertainties based on the distributions and velocities of maser emissions around the nuclei of AGNs. Higher angular-resolution observations of BHEX would allow us to estimate accurate masses of SMBHs in AGN, which enables to calculate the enclosed mass within $\sim$ 0.1\,pc from the nuclei by discriminating the contamination of gas, stars, and dust around the nuclei. 
Accurate masses of SMBHs for megamasers would verify the known strong relationship between the SMBH masses and various properties of their host galaxies.}

\item {Direct measurement of $H_0$: VLBI observations of locations of maser spots and their line-of-sight velocities provide direct geometric distance scales.
Using BHEX the megamasers can be utilized to make the most accurate direct measurement of the Hubble constant in our current Universe by monitoring water masers on rotating disks orbiting around the central SMBHs. The $H_0$ obtained by the megamaers would lead to re-calibration of the Chepheid distance scales used in the HST key project.}

\item 
{Angular momentum transfer in the rotation disk: The angular momentum of the molecular gas in the AGN region plays a key role in the SMBH fueling. It is thought to be linked to the origin of the AGN jet. The 321 GHz water maser requires a physical condition with a higher temperature of $>$ $10^3$ K and a number density of $>$ $10^4$ cm$^{-3}$ compared to that of the 22 GHz maser\cite{Gray_2016}. Thus, the 321 GHz water maser is expected to trace further closer to the SMBH within the innermost radius of the megamaser disk, where the gravitational acceleration is higher and the disk rotation is faster. The combination of 22 and 321 GHz water megamaser studies will provide important clues how the angular momentum of accretion gas is transferred such that the matter can accrete onto a SMBH. }

\end{itemize}

\subsection{Science Opportunities with a Potential Single Dish Mode}
\label{sec:sci-sigle-dish}
The current concept design of BHEX focuses on VLBI operations given the mission-defining science goals\cite{BHEX_Johnson_2024}. 
The mission, launching a single orbital station to space, inherently requires a ground station to conduct a VLBI observation. 
The availability of millimeter/submillimeter ground stations are significantly impacted by numerous factors including the weather conditions, as well as the operational models of each observatory that often has other science operations and regular maintenance seasons.
In fact, the current operation concept assumes VLBI observations every $\sim 3$\,days even for the focused 6-month session on the primary photon ring science\cite{BHEX_Issaoun_2024}, leading to a duty cycle of $\lesssim 30$\,\% for VLBI operations.
Those constraints create a certain large ($\gtrsim 60$\,\%) fraction of the mission lifetime, when the sattelite cannot execute VLBI observations due to the lack of available ground stations.

The BHEX Japan Consortium considers this expected downtime of VLBI operations an unprecedented opportunity to explore the radio sky not accessible from the ground \textit{using BHEX as a standalone observatory}. 
The Japanese radio astronomy community, which historically evolved with millimeter radio astronomy through the Nobeyama 45-m Telescope, the Nobeyama Millimeter Array and now ALMA, has a dominant population of astronomers utilizing molecular emission as a tool to explore the universe. 
Given the background and enthusiasm for extending the ongoing studies of molecular universe to space, the Consortium explored the transformative molecular science that is only possible from space.

With conceptual studies, the Consortium identifies that BHEX will be the {\it highest-sensitivity explorer of oxygen molecules}, a key molecule for understanding the overall taxonomy of the oxygen atom --- the third most abundant element in the universe --- and its role in the chemical evolution of the universe
if BHEX has an additional receiver operating at the 50-70\,GHz band.
To make it possible, BHEX only needs to add an HEMT receiver, which has a well-matured technology for space mission and may be added without a significant cost and power load in the second stage of the cryocooling system (\autoref{sec:cryocooler}). 
The unique nature of the BHEX instrumental system allows that received signals may be digitally sampled in its VLBI digital backend\cite{BHEX_Srinivasan_2024} and further processed on a ground spectrometer after space-to-ground transfer via laser communications\cite{BHEX_Wang_2024}, just like VLBI signals.

Observations at 50-70\,GHz from ground are strictly prohibited due to the severe atomospheric absorption. 
Only with the addition of a dedicated HEMT receiver, BHEX will make the first exploration of the molecular universe in those bands, providing the most sensitive search for oxygen molecules (O$_{2}$) and the first-ever search of complex molecules that fall within this band. In the following, we briefly outline our proposed 'filler' observations executed during the gap of VLBI operations in the BHEX mission. 

\subsubsection{Molecular Oxygen Observations}

The O$_{2}$ molecule has magnetic dipole transitions at millimeter and submillimeter wavelengths. The frequency ranges corresponding to these O$_{2}$ transitions cannot be observed by ground-based telescopes due to their strong atmospheric absorption. Despite extensive efforts by SWAS, Odin and Herschel, interstellar O$_{2}$ lines have rarely been detected\cite{Larsson_2007, Goldsmith_2011}. Some of the O$_{2}$ lines at 50 -- 60 GHz are expected to show maser activities\cite{Bergman_1995}, although no observation has been done in this frequency range. Given the high elemental abundance of oxygen after hydrogen and helium, molecular O$_{2}$ lines would be unique probes for single-dish observations with the BHEX to study the physical and chemical properties of various astronomical objects\cite{Goldsmith_2000}.

In star-forming regions, previous modeling studies suggest that protostellar envelopes with shorter pre-stellar phases ($<$ 0.1~Myr) become molecular oxygen rich, whereas those with longer pre-stellar phases ($>$ 1~Myr) become water rich \cite{Schmalzl_2014, Notsu_2021,vanDishoeck_2021}. Survey observations of molecular oxygen lines with BHEX will clarify the relation between chemical evolution of oxygen bearing molecules and physical conditions (such as timescale of core collapse) of star-forming regions (pre-stellar cores).

Although oxygen chemistry including O$_{2}$ is important to understand the chemical/physical properties of molecular clouds in the nuclear regions of nearby galaxies, the observation of O$_{2}$ is limited to slightly red-shifted galaxies due to the limited atmosphere window. Starburst galaxies will provide new insight into oxygen chemistry because they possess extreme environments, such as extensive PDRs, high cosmic-ray ionization 
rate, and shocks, that are not observed in molecular clouds in the solar neighborhood.  We propose a deep integration observation toward extensively studied nearby starburst galaxy NGC 253 with BHEX. 

\subsubsection{Line Survey}
Protostellar sources are known to show diversity in their chemical compositions\cite{Lefloch_2018}.  Organic molecules, especially those consisting of hydrogen, carbon, and oxygen atoms, have recently been well explored with radio observations.  Meanwhile, chemistry with other common elements (nitrogen, phosphorus, and sulfur atoms) is still far from a thorough understanding.  
The fundamental molecules (HDCO, H$_{2}$COH$^{+}$, D$_{2}$CO, NH$_{2}$OH, HDS, CS$^{+}$, CH$_{2}$D$^{+}$, HCND$^{+}$, etc) exist around 50-60~GHz (e.g., Cologne Database for Molecular Spectroscopy). In addition to the simplest of the configurations, they are in the ground state.

Deep survey observations of their key species with BHEX will provide us with essential information to fill our understanding of the most common element, 'CHNOPS'.
The survey and mapping of these molecules in molecular clouds will lead not only to the field of interstellar chemistry but also to understanding the fundamental physical parameters of star-forming regions.

\section{Potential technical and instrumental contributions}
\label{sec:tech_cases}
BHEX has been designed to be a sensitive broadband instrument\cite{BHEX_Johnson_2024, BHEX_Marrone_2024, BHEX_Peretz_2024}, enabled by a novel sensitive broadband receiver system\cite{BHEX_Tong_2024}, a wideband digital processing system\cite{BHEX_Sridharan_2024} and wideband space-ground downlink through laser communications\cite{BHEX_Wang_2024}.
The instrumental design of BHEX is driven by key science goals of the mission, such as the first direct detection of a photon ring for \m87 and \sgra, and detections of horizon-scale emission from nearby AGN sources.
Key science goals require the fringe sensitivity of $\sim 1$\,mJy\cite{BHEX_Johnson_2024}.
The fringe sensitivity $S_{i,j}$ for a baseline with a pair of two antennas $i$ and $j$ is given by
\begin{align}
S_{i,j} = \frac{\sqrt{{\rm SEFD}_i {\rm SEFD}_j}}{\eta_q \sqrt{2\Delta \nu \Delta t}},
\end{align}
where SEFD is the system equivalent flux density of each antenna, $\Delta \nu$ is the frequency bandwidth, $\Delta t$ is the integration time and $\eta_q$ is an efficiency of VLBI digital processing primarily attributed to the quantization loss. 
The sensitivity requirement results in its current design where (i) the overall VLBI signal chain has a wide bandwidth of 64\,Gbps for larger $\Delta \nu$, (ii) simultaneous dual-band recieving system to utilize the frequency phase transfer technique for longer $\Delta t$, and (iii) the satellite has a sensitive receiver system for lower SEFD.

The BHEX Japan Consortium has identified four key technical and instrumental contributions as a potential cost-sharing partner for the mission. The first two areas are key instruments for its broadband receiver system (\autoref{sec:recsystem}): the 300\,GHz Superconductor-Insulator-Superconductor (SIS) mixer and the cryocooler.
The latter two areas are terrestrial instruments to support the science operation of BHEX (\autoref{sec:groundsupport}): ground-based optical downlink stations and millimeter/submillimeter telescopes. 
As described in the following, all these instruments and technologies are essential for the success of the mission and leveradge the decades of strategic development in Japan on space- and ground-based instruments as well as the key enabling technologies for astronomy and optical communications.

\subsection{Key Instruments for the Receiver System}
\label{sec:recsystem}
As a space antenna for which the diameter of the dish is limited to $\sim$3.5\,m by the size of the payload\cite{BHEX_Johnson_2024,BHEX_Marrone_2024,BHEX_Sridharan_2024}, the sensitivity of BHEX is significantly dependent on the broadband performance of the receiving system. 
The SEFD of a space antenna is defined by
\begin{align}
    {\rm SEFD} &\equiv \frac{2 k T_{\rm R}}{\eta_A A}\\ 
    &\approx 12{,}500\,{\rm Jy} \times \left( \frac{\nu}{100\,{\rm GHz}} \right) \left( \frac{D}{3.5\,{\rm m}} \right)^{-2}  \left( \frac{T_{\rm R}}{30\, {\rm K}} \right) \left( \frac{\eta_{\rm A}}{0.7} \right)^{-1},
\end{align}
where $D$ is the dish effective diameter, $T_{\rm R}$ is the receiver noise temperature, and $\eta_{\rm A}$ is aperture efficiency. The lower bound of $T_{\rm R}$ is given by the fundamental quantum limit of the receiver noise temperature $T_{\rm Q}$ for broadband continuum measurements using a dual-sideband receiver (i.e. $T_Q/T_R < 1$) defined by
\begin{align}
    T_{\rm Q} \equiv \frac{h \nu}{k} &= 4.8\,{\rm K} \times \left( \frac{\nu}{100\,{\rm GHz}} \right).
\end{align}
To maximize sensitivity, the receiver system needs to minimize the receiver noise temperature with respect to the quantum limit across its broad frequency bandwidth.

The BHEX receiving system\cite{BHEX_Tong_2024} is currently designed to have a dual-polarization receiver operating over the 240-320 GHz frequency range, utilizing a SIS mixer, to minimize receiver temperature noise for the 250 and 310\,GHz observations critical for BHEX to achieve the highest angular resolutions ever for key science goals. 
The deployment of a single SIS mixer across two bands was selected by a trade-off study to balance the sensitivity, the overall cost and power consumption tightly constrained for a NASA SMEX mission.
For optimal performance, the dualband 240-320 GHz receiver should operate at a bath temperature of 4.5 K, which necessitates the integration of a cryocooler. 
The BHEX Japan Consortium is exploring potential contributions to supply the SIS mixer for the dual-band 240-320 GHz receiver (\autoref{sec:sismixer}), and a space-qualified cryocooler for the receiving system (\autoref{sec:cryocooler}). 

\subsubsection{SIS Mixer}
\label{sec:sismixer}
\begin{figure}
    \centering
    \includegraphics[width=0.6\textwidth]{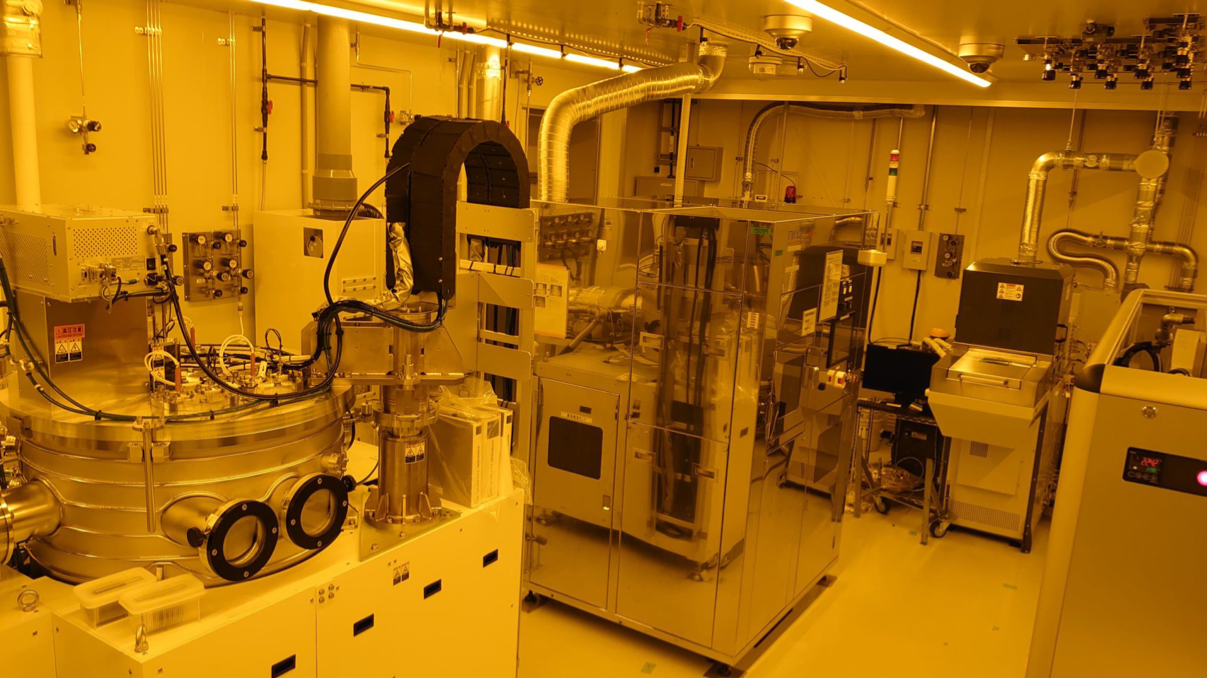}
    \caption{The cleanroom for the SIS junction development at the Advanced Technology Center of National Astronomical Observatory of Japan.}
    \label{fig:naoj-atc-cleanroom}
\end{figure}
In Japan, the Advanced Technology Center (ATC) of the National Astronomical Observatory of Japan has played an important role in the development of SIS mixers. ATC has a cleanroom to develop and produce SIS mixer devices, as shown in \autoref{fig:naoj-atc-cleanroom}. 
In the construction phase of the ALMA telescope, ATC produced all the SIS mixer devices for the ALMA Band 4 (125–163\,GHz), Band 8 (385–500\,GHz), and Band 10 (787–950\,GHz) receivers using standard Nb/Al-AlOx/Nb junctions.\cite{Uzawa_2021}
Nb/Al-AlN/Al/Nb junctions have been currently developing to obtain high critical current densities to increase the operational bandwidth in RF and IF for next-generation receivers to be used in the planned ALMA upgrade in the 2030s\cite{Carpenter_2020} (ALMA2030; also called ALMA2 in Japan), etc. 

ATC has recently successfully demonstrated an SIS mixer that covers 275-500\,GHz using Nb junctions with the current density of 31\,kA/cm$^2$.\cite{Kojima_2018}
As described in Tong\,et\,al.\cite{BHEX_Tong_2024}, BHEX currently plans to deploy a dual-band 240-320 SIS mixer fabricated by NAOJ ATC based on this new SIS mixer. As part of the BHEX Japan Consortium, the receiver group at NAOJ ATC has contributed to the technical concept design studies of the BHEX receiver group.

\subsubsection{Cryocooler}
\label{sec:cryocooler}
The cryocooling system operating at $\sim 4.5$\,K is another critical key component of the BHEX instrument \cite{BHEX_Rana_2024}. %
Japan has been a technology leader for the cryocooling system operating at such cryogenic temperatures for spaceflight --- to date, only three $\sim 4$\,K fully closed-cycle cryocooler systems have ever operated in space, which \textit{all} are for missions led by JAXA and manufactured by Sumitomo Heavy Industries (SHI) in Japan that has been developing cryocoolers for use in space since 1987\cite{Narasaki_2012}.

\begin{figure}
    \centering
    \includegraphics[width=0.45\textwidth]{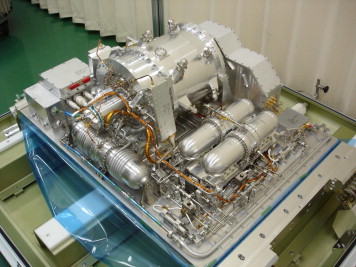}\hspace{3em}
    \includegraphics[width=0.3\textwidth]{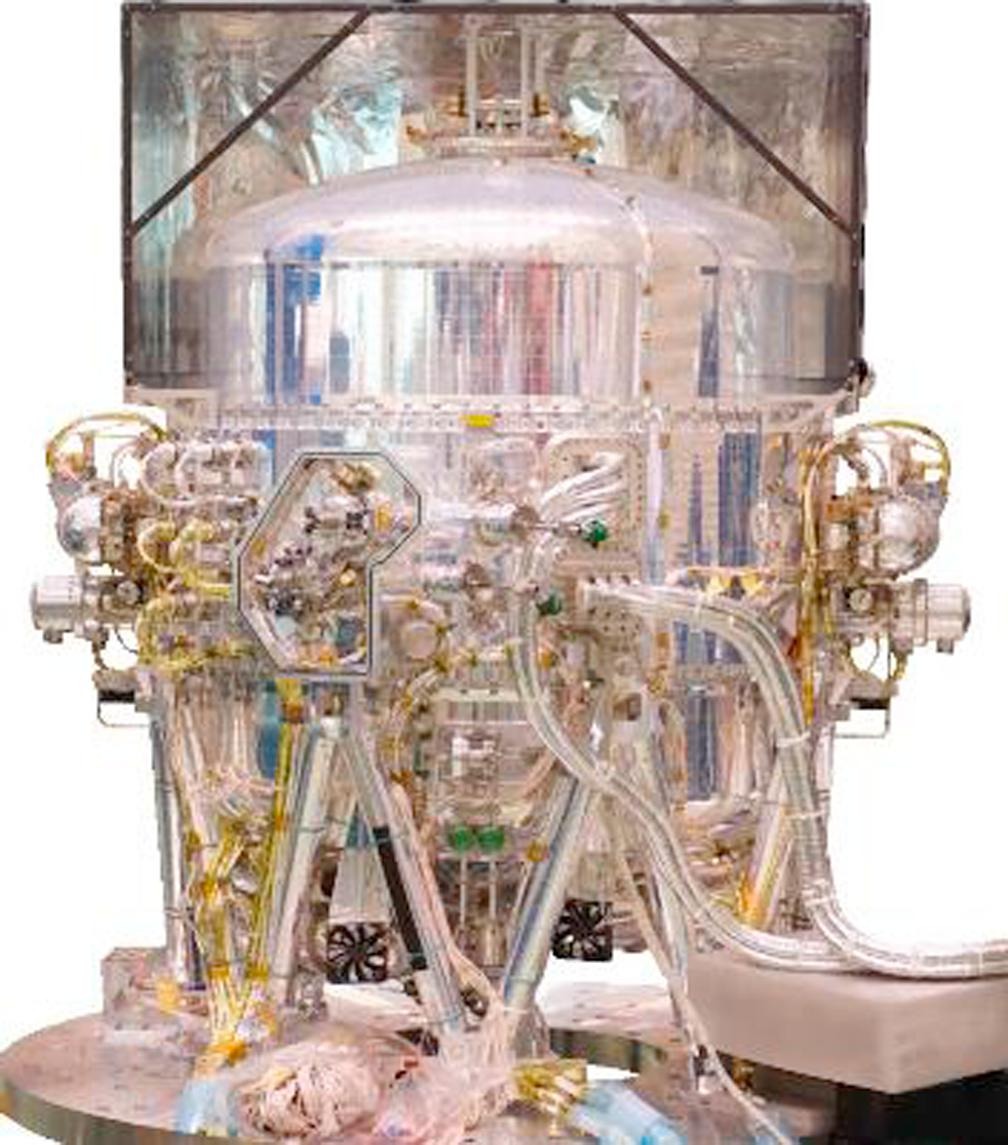}
    \caption{The 4\,K cryocooler systems developed by Sumitomo Heavy Industries and operated in the pioneering NICT-JAXA SMILES mission (left) and the JAXA-NASA Hitomi/XRISM X-ray missions (right). Figures are reprinted from Otsuka et al.\cite{Otsuka_2010} and Yoshida et al.\cite{Yoshida_2018}, respectively, with permission from Elsevier.}
    \label{fig:shi-cryocooler}
\end{figure}

The first space demonstration of a fully closed-cycle cryocooler system 4\,K (\autoref{fig:shi-cryocooler}) was made on the superconducting submillimeter wave limb emission sounder (SMILES) mission \cite{Masuko_1997} on the Japanese Experimental Module (JEM; also called Kibo Module) of the International Space Station (ISS), jointly developed by the National Institute of Information and Communications Technology (NICT) and JAXA. The mission, launched in 2009 to investigate chemical processes related to ozone depletion, used superconductive SIS mixers at the 650\,GHz band and the 4\,K-class mechanical cooler for highly sensitive observation \cite{Ochiai_2010}. 
The 4\,K class fully closed-cycle cryocooler, which consists of a two-stage Stirling cooler and a JT cooler, was manufactured by SHI\cite{Inatani_2005, Otsuka_2010} and operated successfully \cite{Ochiai_2010, Narasaki_2012}. 

With the success of the SMILES/JEM mission, the Hitomi (ASTRO-H) mission, a JAXA-NASA X-ray astronomy space mission, deployed a multi-stage cryocooler developed by SHI for its Soft X-ray Spectrometer (SXS), an X-ray microcalorimeter that requires a thermal bath below 1.3 K\cite{Yoshida_2016} (\autoref{fig:shi-cryocooler}). 
Hitomi, launched in February 2016, unfortunately operated for only a month due to attitude control problems and loss of communication \cite{Takahashi_2018}.
Despite its short lifetime, the observational instruments on board, including the SXS and its cryocooler, consisting of a two-stage Stirling cooler and a JT cooler, functioned as expected \cite{Yoshida_2018, Fujimoto_2018} and successfully delivered astronomical results \cite{2016Natur.535..117H,2017Natur.551..478H,Takahashi_2018}.
The X-Ray Imaging and Spectroscopy Mission (XRISM), a Hitomi recovery mission launched in September 2023 and in operation since then, implemented the microcalorimeter Resolve and associated SHI-manufactured cryocooling system, which both have almost the same design as Hitomi's SXS\cite{Ezoe_2020, Imamura_2023}. This spaceflight cryocooling continues to demonstrate good on-board performance amid spaceflight.

The current BHEX cryocooler system design, described by Rana et al.\cite{BHEX_Rana_2024}, has a cryocooling system with two cold stages: a 20K stage for the HEMT-based 80-106\,GHz receiver and a 4.5K inner stage for the SIS-based 240-320\,GHz receiver. 
The current specifications, including heat loads at each stage and operational temperatures, match well with those accomplished by SHI's cryocoolers in Hitomi and XRISM. 
The BHEX Japan Consortium and the BHEX cryocooler group have explored the possibility of taking advantage of decades of Japanese experience in the space-qualified cryocooling system and using a cryocooler manufactured by SHI for BHEX.

\subsection{Ground Support for the Hybrid Observatory}
\label{sec:groundsupport}
The BHEX mission will operate as a ''hybrid'' observatory, as outlined in Issaoun et al. \cite{BHEX_Issaoun_2024}, that involves a spaceborne radio antenna and ground-based millimeter/submillimeter observatories for VLBI, and a ground-based downlink network that receives the signal from the satellite.
The sensitivity requirement for a broad bandwidth necessitate a downlink infrastructure with optical communications at the bandwidth of $\sim 100$\,Gbps\cite{BHEX_Johnson_2024}, which have demonstrated orders of magnitude higher data transmission rates than radio frequency methods commonly adopted in previous space VLBI missions\cite{BHEX_Wang_2024}.

The ground optical downlink network is a key integral part of BHEX that has an aspect of a broadband instrument at a data recording rate of 64\,Gbps. With more optical ground stations available across the globe, the space telescope can more timely transmit digitally sampled radio signals into the ground, which will allow to increase in the duty cycle of the science operations and potentially reducing the overall power and financial costs for the on-board buffer disk storage. Furthermore, redundancy in the available stations for a given time zone will mitigate the risk of loss or delay in data delivery due to poor weather at some sites around the observation.

As an interferometer, BHEX will benefit from more stations on the ground that will provide better Fourier (or $uv$) coverages, leading to higher image fidelity and a more sensitive imaging sensitivity. 
In particular, the participation of a highly sensitive antenna provides an anchor point for detecting interferometric signals (or fringes) between the BHEX satellite and the ground network. 
Furthermore, ground stations in East Asia uniquely cover the time range where telescopes in Europe and North/South America cannot cover, which is helpful to constrain the intra-day variations of \sgra\cite{EHTC2017SgrAPaper1, EHTC2017SgrAPaper2} as well as other major EHT targets including M87 and other AGN sources which are known to have structural variations on time scales of a few days even at the EHT resolution a few times worse than BHEX \cite{EHTC2017M87Paper3, Kim_2020}.

The BHEX Japan Consortium currently explores supporting its ground operation by providing a ground optical downlink network operated by NICT (\autoref{sec:nictlasercom}) and submillimeter/millimeter radio telescopes operated by NAOJ (\autoref{sec:groundradiostations}).

\subsubsection{Optical Ground Station Network in Japan}
\label{sec:nictlasercom}

\begin{figure}
  \centering
  \begin{subfigure}{0.45\textwidth}
    \includegraphics[width=\linewidth]{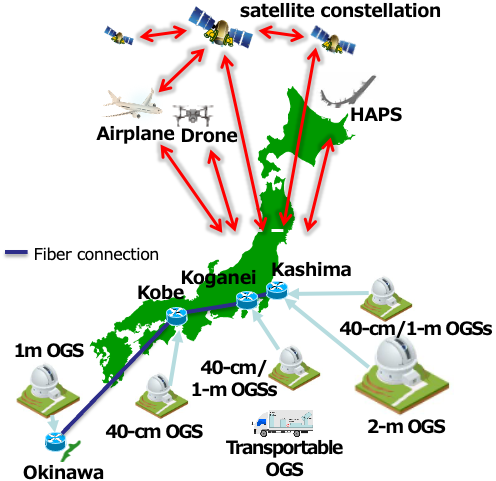}
    \caption{}
  \end{subfigure}
  \hspace{1em}
  \begin{subfigure}{0.5\textwidth}
    \includegraphics[height=0.25\textheight]{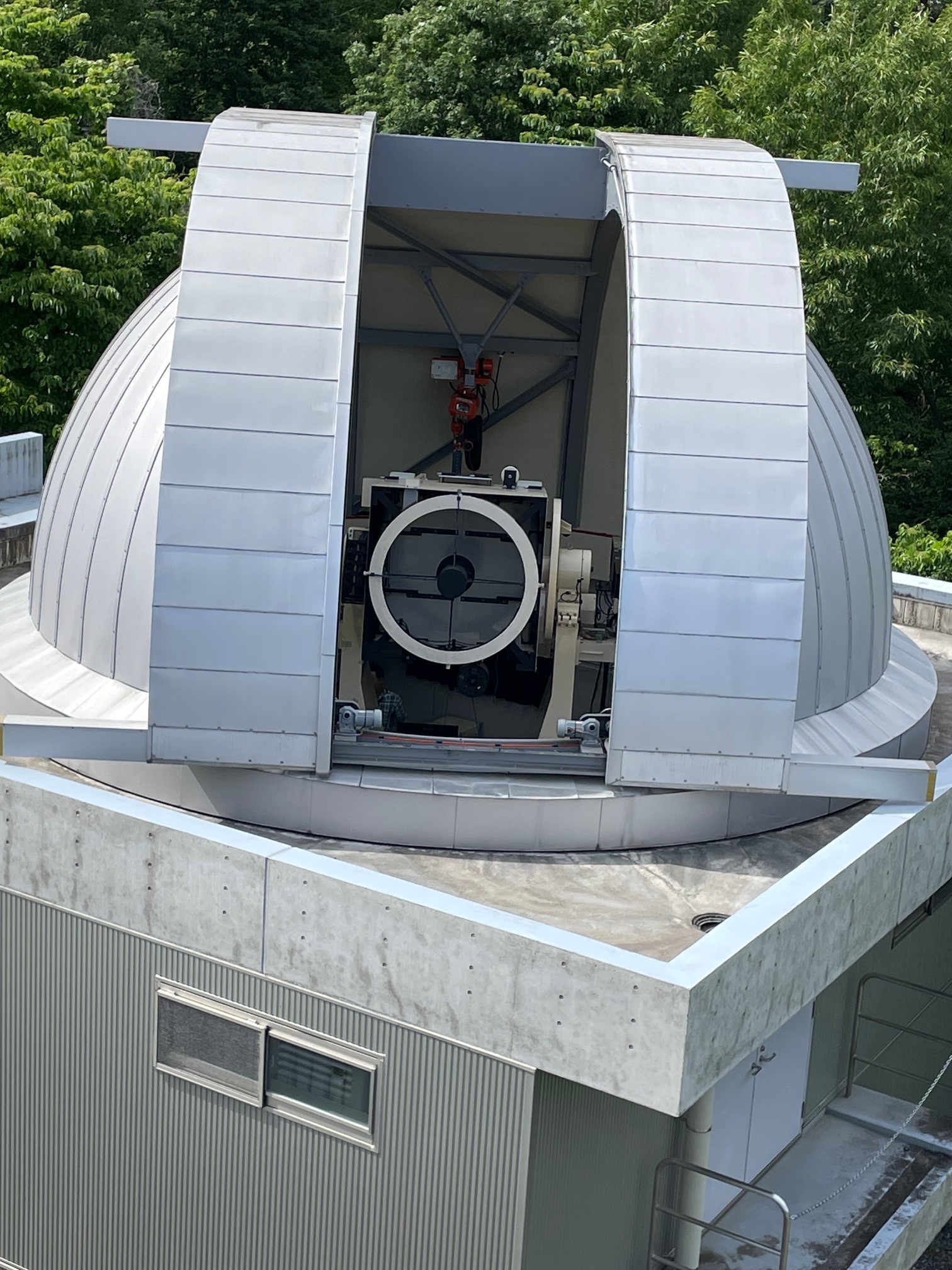}\hspace{0.3em}\includegraphics[height=0.25\textheight]{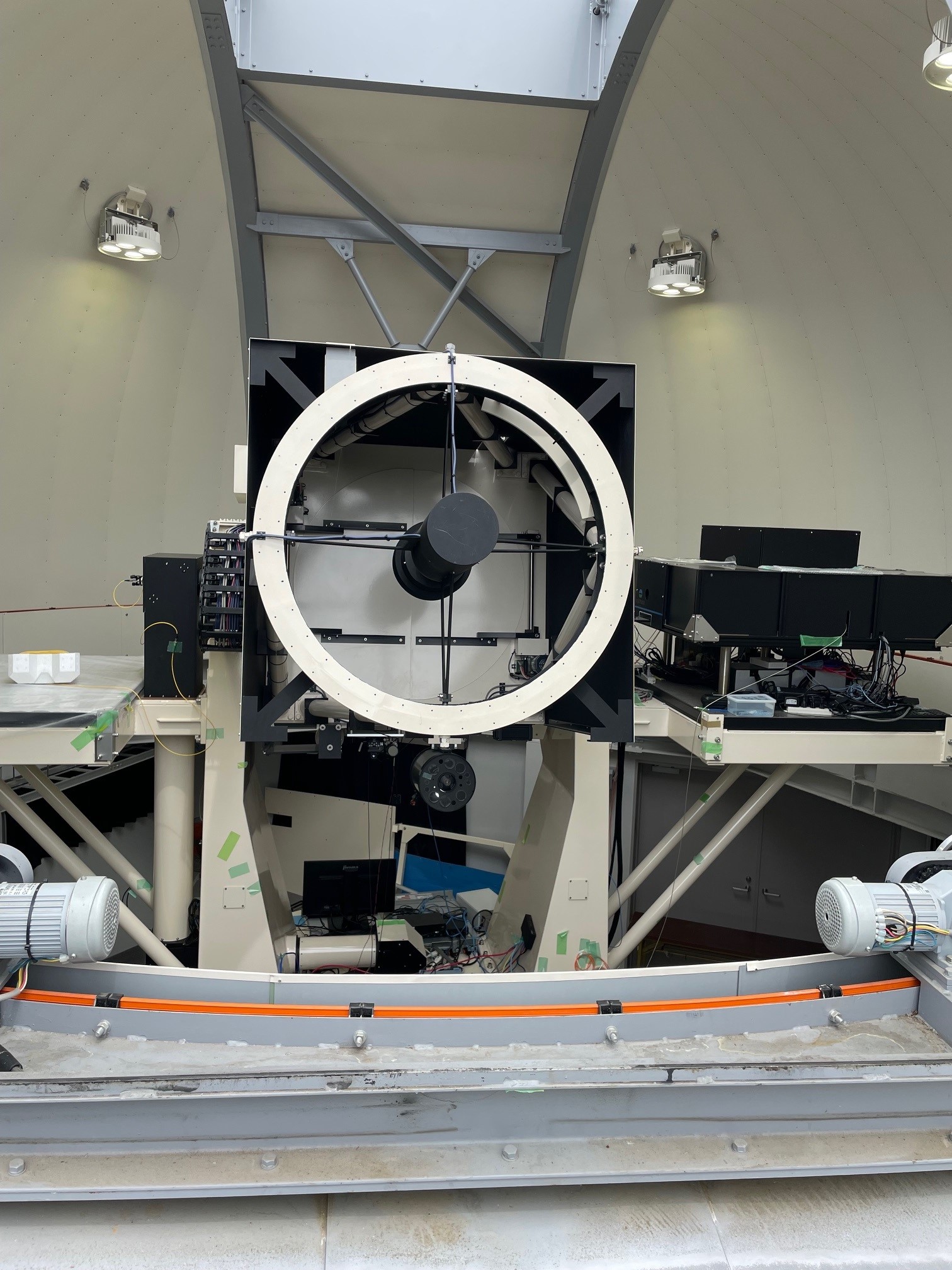}
    \caption{}
  \end{subfigure}
  \caption{The optical ground stations in Japan, which has been developed by the NICT Space Communication Systems Laboratory. (a) A schematic view of the network. Figure credit: Dimitar Kolev. (b) The Koganei 1-m Telescope at the NICT headquaters in Koganei, Tokyo. A prototype adaptive optics system is implemented on the optical bench at the left nasmyth table\cite{Kolev_2019}. Figure credit: Hideki Takami.}
    \label{fig:nict-ogs}
\end{figure}

In Japan, the NICT Space Communication Systems Laboratory (SCSL) has been playing a central role in the development and implementation of the nation-wide network of optical ground stations (OGSs) for the use of big data in space enabled by optical ground-space laser communications. 
The OGS network, currently under development, involves five stations with three different diameters at four geographic sites that span from eastern Japan to the southern-west end (\autoref{fig:nict-ogs}), as summarized below.
\begin{itemize}
\item A 2-m telescope equiping an adaptive optics (AO) system at the NICT Kashima Space Technology Center in Ibaraki, now under constructions, for deep-space communcation and 100\,Gbps class laser communications.
\item Three 1-m telescopes, each located at the NICT headquaters at Koganei in Tokyo, at Kashima, and at the NICT Electromagnetic Technology Center in Okinawa, for optical communications from geosyncronous orbits (GEO) to ground. The Koganei 1-m telescope (\autoref{fig:nict-ogs}), which serves as a development hub for the OGS instruments, has already integrated a prototype AO system, currently using the central 40 cm of the aperture, for testing and system verifications\cite{Kolev_2019}. NICT plans to extend the AO system to the remaining sites.
\item Three 40-cm telescopes each located at Koganei, Kashima and the NICT Advanced ICT Research Institute in Kobe, Hyogo for space-to-ground link from low Earth orbits (LEO). In addition, to improve connectivity, NICT is developing a portable OGS equipped with a 40-cm telescope.
\end{itemize}
The four locations, equipped with diverse OGS telescopes and with redundancies between sites, are spread over a maximum distance of approximately $\sim 1,600$\,km. 
This arrangement increases the operational reliability of the OGSs in the face of possible poor weather conditions at certain sites. 
The OGS network is interconnected through a high-speed optical fiber network between sites for the transfer of the received data, ensuring uninterrupted space-to-ground communication despite potential variations in the receiving station.

The current baseline specification of BHEX requires each OGS to have (a) a small optical telescope with a diameter approximatively between 0.7 and 1\,m without an AO system, or (b) an AO-equiped telescope with a diameter approximately from 1 to 3\,m, to achieve the sensitivity required to establish the link with the BHEX sattelite\cite{BHEX_Issaoun_2024, BHEX_Wang_2024}. 
Four NICT OGS located in three locations (Kashima, Koganei, and Okinawa) meet this crucial requirement for OGS stations.
In addition, this network of the BHEX-elgible OGS stations has a maximum baseline length of 1, 600\,km, allowing full leberading of the resiliency of the NICT network against adverse weather conditions.
The NICT OGS network may provide a redundant set of stations at a location uniquely covering a wide-range of the sky seen from East Asia for BHEX, which would secure the time coverage for the spact-to-ground data transfer and enhacing the duty cycle of its science operations. 
Within the BHEX Japan Consortium, NICT SCSL has been investigating the potential use of their OGS network for BHEX, as part of the mission-wide concept design studies for the BHEX laser communication and ground operations.

\subsubsection{Ground Millimeter/Submillimeter Observatories}
\label{sec:groundradiostations}
Japan has played an important role in the development of the current EHT array. Beginning with the initial VLBI experiment at 230\,GHz using the Japanese ALMA pathfinder, the Atacama Submillimeter Telescope Experiment (ASTE) 11-m Telescope, in the early 2010s, Japan has been part of the ALMA Phasing Project \cite{Matthews_2018}, integrating essential VLBI features into the ALMA 12-m array, a crucial component of both the EHT and GMVA arrays. As a principal international partner sharing costs for ALMA, NAOJ has spearheaded global collaboration in East Asia. Japan also supports the operations of the James Clerks Maxwell Telescope (JCMT), another EHT facility, through the East Asian Observatory\cite{Asada_2017, EHTC2017M87Paper2}. 
Both ALMA and JCMT are important terrestrial stations for BHEX science operations \cite{Issaoun_2022}.

\begin{figure}
  \centering
  \begin{subfigure}{.4\textwidth}
    \centering
    \includegraphics[height=0.13\textheight]{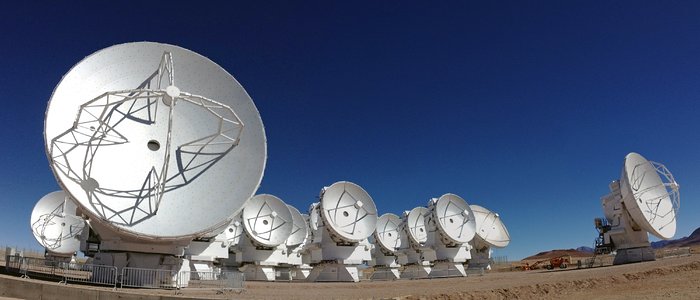}
    \caption{TPA in ALMA ACA}
  \end{subfigure}%
  \begin{subfigure}{.3\textwidth}
    \centering
    \includegraphics[height=0.18\textheight]{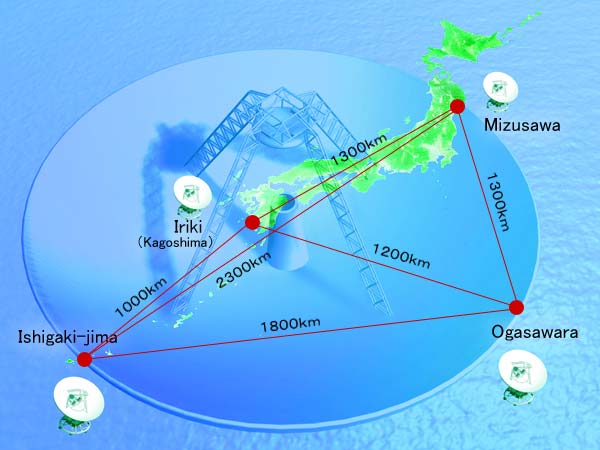}
    \caption{VERA}
  \end{subfigure}%
  \begin{subfigure}{.3\textwidth}
    \centering
    \includegraphics[height=0.18\textheight]{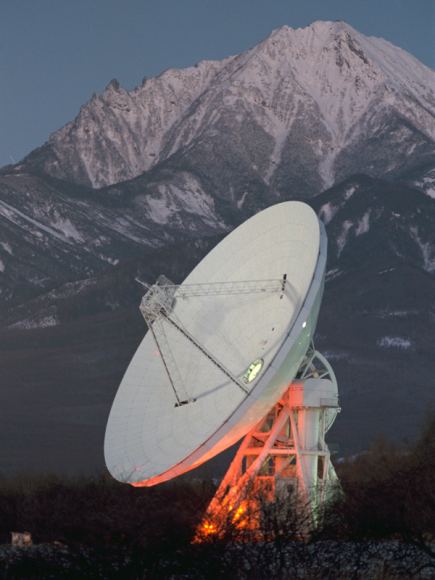}
    \caption{Nobeyama 45-m Telescope}
  \end{subfigure}
  \caption{Three millimeter/submillemeter facilities developed by NAOJ, which can potentially be used to ground support for BHEX, alongside two existing EHT stations (the ALMA main array and JCMT) supported by Japan. (a) Total Power Array in ALMA ACA. The TPA is an ATA subarray with four 12 m antennas located at the four outer edges of the ACA that surround the remaining twelve 7 m antennas. Figure credit: ALMA (ESO/NAOJ/NRAO). (b) VERA to be extended to support 86\,GHz VLBI operation. Figure credit: NAOJ. (c) Nobeyama 45-m telescope, the most sensitive 86 GHz VLBI station in East Asia. Figure credit: NAOJ.}
  \label{fig:groundstations}
\end{figure}

Besides the two current EHT sites, the BHEX Japan Consortium is exploring the use of three additional millimeter/submillimeter facilities developed by NAOJ (\autoref{fig:groundstations}), as outlined below. These facilities have been established in alignment with the priorities of the VLBI and the broader radio astronomy community in Japan for the 2030s.

\paragraph{ALMA Total Power Array (TPA):} ALMA TPA is a subarray of the Atacama Compact Array (ACA; also called ''Morita Array''). 
Among the 66 ALMA antennas, the ACA utilizes twelve 7-m antennas and four 12-m antennas, all built by NAOJ, the latter of which forms TPA. As part of the ALMA 2030 roadmap, the Korean Astronomy and Space Science Institute (KASI), in collaboration with NAOJ, leads the development of the GPU-based spectrometer \cite{Carpenter_2020}.
The TPA is designed to work as an independent subarray to measure the absolute brightness of the targets, whose digitized data will then be processed at the spectrometer and later combined with data from the rest 7-m ACA array and/or the main 12-m array of ALMA.
In addition to the main ALMA array, which is already a part of EHT and GMVA \cite{Matthews_2018, EHTC2017M87Paper2}, the TPA, which can observe both 100 and 300\,GHz bands, may offer a sensitive, potentially less subscribed anchor for BHEX.

\paragraph{VERA:} VERA is a Japanese VLBI array compising four 20-m telescopes (Iriki, Ishigakijima, Mizusawa, and Ogasawara) \cite{Honma_2000}. VERA, primarily designed for the precice VLBI astrometry of Galactic maser sources to study the structures of the Milkey Way galaxy, has been operating primarily at 6.7\,GHz, 22\,GHz and 43\,GHz targetting Methanol, Water and SiO masers. Each antenna of VERA was designed for the potential extension to 86\,GHz\cite{Honma_2000}. With the advent of stations operating at 86\,GHz in East Asia\cite{Akiyama_2022}, a new dual-polarization cryogenic 86\,GHz receiver is being developed for VERA to significantly expand the 86\,GHz network of EAVN \cite{Kameyama_2024}. VERA is anticipated to be available as a ground network for BHEX observations at the 100\,GHz band.

\paragraph{Nobeyama 45\,m Telescope:} Nobeyama 45\,m Telescope is one of the largest millimeter telescopes to date operating up to $\sim 100\,$GHz, known as a historic station of global 86\,GHz VLBI arrays since 1988 \cite{Baath_1991,Baath_1992}. The Nobeyama 45\,m telescope has been part of JVN and EAVN. 
Its receiving system was recently upgraded to support simultaneous tri-band observations at 22, 43 and 86\,GHz for both the single dish and VLBI modes under the Hybrid Integration Project in Nobeyama, Triple-band Oriented (HINOTORI) \cite{Okada_2020, Tsutsumi_2023, Imai_2023}. 
Nobeyama 45\,m Telescope is currently the most sensitive 86\,GHz VLBI station in East Asia, with a strong potential to be a sensitive anchor station for BHEX observations at the 100\,GHz band.

\section{Summary and Outlook}
\label{sec:summary}
We have described our current vision for the BHEX mission. 
The BHEX Japan Consortium has actively participated in the ongoing development of the mission concept. 
The Consortium has identified a broad spectrum of scientific inquiries that could be pursued through the mission, driven by ongoing research in Japan. This spans from horizon-scale black hole astrophysics to the studies of varied AGN populations. 
The science vision even extends to the molecular universe, which could be investigated for the first time if a specialized HEMT receiver is integrated into the satellite. 
Furthermore, the Consortium has delineated four principal domains of technological and instrumental contributions, capitalizing on the community's long-standing expertise and strategic initiatives: these include the broadband 300\,GHz SIS mixer and the 4.5K cryocooler system for the BHEX satellite, along with terrestrial networks of OGS and radio telescopes to bolster BHEX ground operations. 

After nearly a year of studies in Japan, the BHEX mission has emerged as an unprecedented scientific opportunity for the Japanese community to deepen our knowledge of the universe. This includes addressing fundamental questions about black holes and their cosmic context, along with the roles of oxygen molecules within the molecular universe. Concurrently, the mission offers a prime opportunity to employ and enhance the technologies that have been strategically cultivated over the years in various fields, apart from astronomy.

Given the strong benefits of Japan joining the BHEX mission as an official partner, the BHEX Japan Consortium now aims to develop the mission as an official JAXA project.
The envisioned Japanese contributions to BHEX would fit well into the scope of a JAXA's framework for international space missions, named the strategic international collaboration program. 
The program is designed for Japan to participate in a foreign space mission as a minor cost-sharing partner optionally with a mission critical instrument strategically developed in Japan. 
In June 2024, the Consortium will host the Black Hole Explorer Japan Workshop\footnote{\url{https://sites.mit.edu/bhex-japan-workshop-2024/} accessed on May 29, 2024} in NAOJ, assembling more than a hundred scientists both domestically and internationally to further promote the engagmenet of the broad Japanese communties into the international BHEX communities. 
In tight collaboration with the broader BHEX communities, the Consortium aims to further deepen and accelerate the ongoing mission concept development in Japan, with the planned formation of an official Working Group under the JAXA's Space Science Committee.

\acknowledgments %
\input{acknowledgement}

\end{document}

%% file: authorlist.tex
\newcommand{\mithaystack}{Massachusetts Institute of Technology Haystack Observatory, Westford, MA 01886, USA}
\newcommand{\naojmizusawa}{Mizusawa VLBI Observatory, National Astronomical Observatory of Japan, Iwate 023-0861, Japan}
\newcommand{\harvardbhi}{Black Hole Initiative at Harvard University, Cambridge, MA 02138, USA}
\newcommand{\yamaguchi}{Graduate School of Sciences and Technology for Innovation, Yamaguchi University, Yamaguchi 753-8512, Japan}
\newcommand{\isasjaxa}{Institute of Space and Astronautical Science, Japan Aerospace Exploration Agency, Kanagawa 252-5210, Japan}
\newcommand{\sokendaijaxa}{Department of Space and Astronautical Science, The Graduate University for Advanced Studies, SOKENDAI, Kanagawa, 229-8510, Japan}
\newcommand{\harvard}{Harvard University, Cambridge, MA 02138, USA}
\newcommand{\toyo}{Natural Science Laboratory, Toyo University, Tokyo 112-8606, Japan}
\newcommand{\tokyodenki}{Division of Science, School of Science and Engineering, Tokyo Denki University, Saitama 350-0394, Japan}
\newcommand{\cfa}{Center for Astrophysics $|$ Harvard \& Smithsonian, Cambridge, MA 02138, USA}
\newcommand{\utokyoicrr}{Institute for Cosmic Ray Research, The University of Tokyo, Chiba 277-8582, Japan}
\newcommand{\utokyoastro}{Department of Astronomy, Graduate School of Science, The University of Tokyo, Tokyo 113-0033, Japan}
\newcommand{\niigata}{Graduate School of Science and Technology, Niigata University, Niigata 950-2181, Japan}
\newcommand{\naojatc}{Advanced Technology Center, National Astronomical Observatory of Japan, Tokyo 181-8588, Japan}
\newcommand{\utsukuba}{Center for Computational Sciences, University of Tsukuba, Ibaraki 305-8577, Japan}
\newcommand{\gifu}{Faculty of Engineering, Gifu University, Gifu 501-1193, Japan}
\newcommand{\nict}{Space Communication Systems Laboratory, Wireless Networks Research Center, National Institute of Information and Communications Technology, Tokyo 184-8795, Japan}
\newcommand{\ncu}{The Graduate School of Science, Nagoya City University, Aichi 467-0001, Japan}
\newcommand{\naojmitaka}{National Astronomical Observatory of Japan, Tokyo 181-8588, Japan}
\newcommand{\azabu}{Azabu Junior and Senior High School, Tokyo 106-0046, Japan}
\newcommand{\asiaa}{Institute of Astronomy and Astrophysics, Academia Sinica, Taipei 10617, Taiwan, R.O.C.}
\newcommand{\kogakuin}{Kogakuin University of Technology \& Engineering, Academic Support Center, 2665-1 Nakano, Hachioji, Tokyo 192-0015, Japan}
\newcommand{\tdli}{Tsung-Dao Lee Institute, Shanghai Jiao Tong University, Shanghai 201210, People's Republic of China}
\newcommand{\sjtu}{School of Physics and Astronomy, Shanghai Jiao Tong University, Shanghai 200240, People's Republic of China}
\newcommand{\iaacsic}{Instituto de Astrofísica de Andalucía-CSIC, E-18008 Granada, Spain}
\newcommand{\sokendainaoj}{Astronomical Science Program, The Graduate University for Advanced Studies, SOKENDAI, Tokyo 181-8588, Japan}
\newcommand{\utokyoeaps}{Department of Earth and Planetary Science, Graduate School of Science, The University of Tokyo, Tokyo 113-0033, Japan}
\newcommand{\kyotouyukawa}{Center for Gravitational Physics, Yukawa Institute for Theoretical Physics, Kyoto University, Oiwake-cho, Kitashirakawa, Sakyo-ku, Kyoto-shi, Kyoto-fu 606-8502, Japan}%
\newcommand{\tokyotech}{Institute of Innovative Research, Tokyo Institute of Technology, Tokyo 152-8551, Japan}
\newcommand{\omu}{Graduate School of Science, Osaka Metropolitan University, 1-1 Gakuen-cho, Naka-ku, Sakai, Osaka 599-8531, Japan}
\newcommand{\nittokyo}{National Institute of Technology, Tokyo College, Tokyo 193-0997, Japan}
\newcommand{\kyoto}{Department of Astronomy, Graduate School of Science, Kyoto University, Kyoto 606-8502, Japan}
\newcommand{\sit}{Materials Science and Engineering, College of Engineering, Shibaura Institute of Technology, Tokyo 135-8548, Japan}
\newcommand{\daido}{Daido University, Aichi 457-0819, Japan}
\newcommand{\ipmu}{Kavli Institute for the Physics and Mathematics of the Universe (WPI), University of Tokyo Institutes for Advanced Study, University of Tokyo, Chiba 277-8583, Japan}
\newcommand{\riken}{Star and Planet Formation Laboratory, RIKEN Cluster for Pioneering Research, Saitama 351-0198, Japan}

\author[1,2,3]{Kazunori Akiyama}
    \affil[1]{\mithaystack}
    \affil[2]{\naojmizusawa}
    \affil[3]{\harvardbhi}
\author[4]{Kotaro Niinuma}
    \affil[4]{\yamaguchi}
\author[5,2]{Kazuhiro Hada}
    \affil[5]{\ncu}
\author[6,7]{Akihiro Doi}
    \affil[6]{\isasjaxa}
    \affil[7]{\sokendaijaxa}
\author[8]{Yoshiaki Hagiwara}
    \affil[8]{\toyo}
\author[9]{Aya E. Higuchi}
    \affil[9]{\tokyodenki}
\author[2,10]{Mareki Honma}
    \affil[10]{\utokyoastro}
\author[11]{Tomohisa Kawashima}
    \affil[11]{\utokyoicrr}
\author[12]{Dimitar Kolev}
    \affil[12]{\nict}
\author[13]{Shoko Koyama}
    \affil[13]{\niigata}
\author[14]{Sho Masui}
    \affil[14]{\naojatc}
\author[15]{Ken Ohsuga}
    \affil[15]{\utsukuba}
\author[16]{Hidetoshi Sano}
    \affil[16]{\gifu}
\author[12]{Hideki Takami}
\author[3,15]{Yuh Tsunetoe}
\author[14]{Yoshinori Uzawa}
\author[2]{Takuya Akahori}
\author[4]{Yuto Akiyama}
\author[14, 3]{Peter Galison}
    \affil[14]{\harvard}
\author[2, 15]{Takayuki J. Hayashi}
    \affil[15]{\azabu}
\author[2, 16]{Tomoya Hirota}
    \affil[16]{\sokendainaoj}
\author[17]{Makoto Inoue}
    \affil[17]{\asiaa}
\author[2]{Yuhei Iwata}
\author[18, 3]{Michael D. Johnson}
    \affil[18]{\cfa}
\author[19, 2]{Motoki Kino}
    \affil[19]{\kogakuin}
\author[10,2]{Yutaro Kofuji}
\author[20, 21]{Yosuke Mizuno}
    \affil[20]{\tdli}
    \affil[21]{\sjtu}
\author[22]{Kotaro Moriyama}
    \affil[22]{\iaacsic}
\author[23,16]{Hiroshi Nagai}
    \affil[23]{\naojmitaka}
\author[10,2]{Kenta Nakamura}
\author[24, 10, 25]{Shota Notsu}
    \affil[24]{\utokyoeaps}
    \affil[25]{\riken}
\author[12]{Fumie Ono}
\author[26]{Yoko Oya}
    \affil[26]{\kyotouyukawa}
\author[2]{Tomoaki Oyama}
\author[18,3]{Hannah Rana}
\author[27]{Hiromi Saida}
    \affil[27]{\daido}
\author[4, 28]{Ryo Saito}
    \affil[28]{\ipmu}
\author[12]{Yoshihiko Saito}
\author[29,2]{Mahito Sasada}
    \affil[29]{\tokyotech}
\author[30]{Satoko Sawada-Satoh}
    \affil[30]{\omu}
\author[31]{Mikiya M. Takahashi}
    \affil[31]{\nittokyo}
\author[10,2]{Mieko Takamura}
\author[17]{Edward Tong}
\author[12]{Hiroyuki Tsuji}
\author[32]{Shogo Yoshioka}
    \affil[32]{\kyoto}
\author[33]{Yoshimasa Watanabe}
    \affil[33]{\sit}

%% file: acknowledgement.tex
The mission concept studies for BHEX within the BHEX Japan Consortium have been financially supported by the following programs and organizations.
\begin{itemize}
\item MEXT/JSPS Grants-in-Aid for Scientific Research (KAKENHI) Grant Numbers: JP24684011 (T.H.), JP15H00784 (K.N.), JP15H03644 (Y.H.), JP19KK0081 (M.H), JP17K05398 (T.H.), JP18KK0090 (K.H.), JP18H03721 (K.N.), JP19H01943 (K.H., Y.H., K.N.), JP21H04488 (K.O.), JP21K03628 (S.S.-S., K.N.),  JP21K13954 (Y.O.), JP22H00157 (K.H., S.K., K.N., M.K.), JP22H04955 (Y.U.), JP23K13155 (S.N.), JP23KJ0329 (S.N.), JP23H00117 (T.K.), JP23H00118 (K.N.), JP23K03448 (T.K.), JP23K03453 (S.K.)
\item MEXT as “Program
for Promoting Researches on the Supercomputer
Fugaku” (Structure and Evolution of the Universe Unraveled
by Fusion of Simulation and AI; Grant Number
JPMXP1020240219 (K.O., T.K.) / Black hole accretion disks and quasi-periodic oscillations revealed by general relativistic hydrodynamics simulations and general relativistic radiation transfer calculations; Grant Number JPMXP1020240054 (K.O., T.K.) 
\item the Joint Institute for Computational Fundamental Science (JICFuS; K.O.)
\item the Multidisciplinary Cooperative Research Program in CCS, University of Tsukuba
(K.O.)
\item the ULVAC-Hayashi Seed Fund from the MIT-Japan Program at MIT International Science and Technology Initiatives (MISTI)
\item grants from the National Science Foundation (NSF; AST-1935980, AST-2034306)
\end{itemize}
In addition to the above programs, the BHEX mission concept studies have been supported by
the Smithsonian Astrophysical Observatory,
the Internal Research and Development (IRAD) program at NASA Goddard Space Flight Center,
the University of Arizona,
and
the Black Hole Initiative at Harvard University. 
The authors acknowledge financial supports by
NSF (AST-2107681, AST-2132700, AST-2307887, OMA-2029670), 
the Gordon and Betty Moore Foundation (GBMF-10423), 
and
the Brinson Foundation.
The Black Hole Initiative at Harvard University is funded by
grants from the John Templeton Foundation and the Gordon and Betty Moore Foundation to Harvard University. 
BHEX is funded in part by generous support from Mr. Michael Tuteur and Amy Tuteur, MD. 
BHEX is supported by initial funding from Fred Ehrsam.